\documentstyle[twoside, fleqn,espcrc2]{article}

\def\stackrel#1#2{\mathrel{\mathop{#2}\limits^{#1}}}

\title{Functional Approach to Classical Yang-Mills Theories}

\author{P. Carta\address{Dipartimento di Fisica, Universit\`a di Cagliari and INFN, Sezione di Cagliari,
Cittadella Universitaria, 09042 Monserrato, Italy}
and D. Mauro\address{Dipartimento di Fisica Teorica, Universit\`a di Trieste and INFN, Sezione di Trieste,
Strada Costiera 11, I-34014 Trieste, Italy}}

\begin{document}
\begin{abstract}
Sometime ago it was shown that the operatorial approach to classical mechanics, pioneered in the 30's
by Koopman and von Neumann, can have a functional version. In this talk we will extend this functional approach
to the case of classical field theories and in particular to the Yang-Mills ones. We shall show that the issues of
gauge-fixing and Faddeev-Popov determinant arise also in this classical formalism.
\end{abstract}
\maketitle
\section{Introduction}
In this talk we want to apply the formalism of the Classical Path Integral (CPI) 
\cite{Gozzi1} to a classical field theory with non-abelian gauge invariance. 
The reasons to embark on this work are basically two. The first is geometrical.
As it has been explained in a paper by E. Gozzi in these proceedings \cite{Gozzi2}, 
the CPI provides several geometrical tools like forms, exterior derivatives,
Lie-derivatives etc. These tools may help us in probing the geometrical aspects of Yang-Mills (YM) 
theories which play an important role
in the study of anomalies and similar phenomena. The second reason to spend time on the classical
aspects of field theories is because the classical solutions are the basic ingredients of non-perturbative
approaches to quantum field theory which are needed to study phenomena
like confinement, mass spectrum calculations and so on.
The gauge invariance of YM theories creates several problems in the
implementation of the CPI procedure; so we prefer to start with the 
simple example of a scalar field $\varphi$
with a $\varphi^4$ interaction:
\begin{equation}
{\cal{L}}=\frac{1}{2}(\partial_{\mu}\varphi)(\partial^{\mu}\varphi)-\frac{1}{2}m^2\varphi^2-\frac{1}{24}g\varphi^4
\end{equation}
Calling $\pi$ the momentum conjugate to $\varphi$ we can easily perform the Legendre transform and construct
the associated Hamiltonian:
\begin{equation}
H=\int d^3x\biggl[\frac{1}{2}\pi^2+\frac{1}{2}(\partial_r \varphi)^2+
\frac{1}{2}m^2\varphi^2+\frac{1}{24}g\varphi^{4}\biggr]
\end{equation}
The equations of motion can be written in the following form:
$\displaystyle
\dot{\varphi}(x)=\frac{\delta H}{\delta \pi(x)},\;\dot{\pi}(x)=-\frac{\delta
H}{\delta \varphi(x)}$.
We note a first formal difference between the classical mechanics of a point particle and the classical field theory:
the presence of functional derivatives in the equations of motion. 
Calling the field and its conjugate momentum as $\xi^a=(\varphi,\pi)$, 
with $a=1,2$ and introducing a $2\times 2$ antisymmetric matrix $\omega^{ab}$
we can write the equations of motion in the following compact form:
$\displaystyle \dot{\xi}^a=\omega^{ab}\frac{\delta H}{\delta \xi^{b}}$.
We can then go along the usual steps of the CPI procedure \cite{Gozzi2} 
in order to have an exponential weight in the
path integral:
\begin{equation}
\displaystyle Z^{\varphi^4}_{\scriptscriptstyle CPI}=
\int {\cal{D}}\xi^a{\cal{D}}\lambda_a{\cal{D}}c^a{\cal{D}}\bar{c}_a \;e^{i\int d^4x
\tilde{\cal{L}}^{\varphi^4}} \label{pippo}
\end{equation}
where the {\it Lagrangian density} has the same form as that associated to the CPI of the point particle:
\begin{eqnarray}
\displaystyle \widetilde{\cal L}^{\varphi^{4}}&=&\lambda_a(\dot{\xi}^a-
\omega^{ab}\frac{\delta H}{\delta \xi^b})\nonumber\\
&&+i\bar{c}_a(\partial_t\delta^a_b-
\omega^{ac}\frac{\delta}{\delta \xi^c}\frac{\delta}{\delta \xi^b}H)c^b
\end{eqnarray}
What are then the main features that make the CPI of a 
point particle different from the CPI of a field theory? The fact that a
field has $\infty$ degrees of freedom (one for every point of the space $\vec{x}$)
has several consequences:
{\bf a)} we have to label the fields with an $\vec{x}$;
{\bf b)} the Lagrangian density $\widetilde{\cal L}$ in (\ref{pippo})
is integrated over $d^4x$ and not only over the time
variable; 
{\bf c)} in the functional measure ${\cal{D}}\xi^{a}=\prod_{k\;\vec{x}}
d\xi^{a}(k,\vec{x})$ the product is extended also to space variables.

Having shown the basic formal aspects of a simple CPI field theory, we can now go on with the 
case of Yang-Mills theories. The main 
ingredients are:
{\bf 1)} the fields $A^{\mu}_a(x)$ with $a$ colour index running from 1 to $n$, being $n$ the
number of generators $T_a$ of a Lie algebra;
{\bf 2)} the structure constants $C^{bc}_a$ appearing in the algebra of the $T_a$: $[T_a,T_b]=iC_{ab}^cT_c$;
{\bf 3)} the antisymmetric tensor $F_a^{\mu\nu}=
\partial^{\mu}A^{\nu}_a-\partial^{\nu}A^{\mu}_a-gC_a^{bc}A^{\mu}_bA^{\nu}_c$
which enters into the Lagrangian density in the usual way:
${\cal{L}}=-\frac{1}{4}F_a^{\mu\nu}F_{\mu\nu}^a$.

\section{A natural gauge-fixing procedure}

If we want to construct, as in the case of a $\varphi^4$ theory, 
the associated Hamiltonian we have to solve the problem of the constraints.
The {\it primary} ones derive from the definition itself of conjugate momenta: 
$\displaystyle \pi^{\scriptscriptstyle 0}_a=\frac{\partial\cal L}{\partial\dot{A}^a_{\scriptscriptstyle 0}}=0$. 
The {\it secondary} ones
$\sigma_a=-\vec{\nabla}\cdot\vec{\pi}_a+C^b_{\;ac}\pi^k_bA^c_k=0$ derive instead from the requirement
that the primary constraints have to be conserved under time evolution.
Because of these constraints we cannot write the Hamiltonian in a {\it unique} way.  In fact,
following \cite{Henneaux1}, we can write the Hamiltonian as:
\begin{equation}
\displaystyle
H=\int\, d^3x\Bigl(\frac{1}{2}\vec{\pi}_a\cdot\vec{\pi}^a
+\frac{1}{2}\vec{B}_a\cdot{\vec{B}}^a+\lambda^a\sigma_a\Bigr) \label{ma}
\end{equation}
It is given by the sum of the YM Hamiltonian 
$\displaystyle {\cal H}_{\scriptscriptstyle YM}=\frac{1}{2}\vec{\pi}_a\cdot\vec{\pi}^a+\frac{1}{2}\vec{B}_a\cdot
\vec{B}^a$ and an arbitrary combination of the secondary constraints $\lambda^a\sigma_a$
where
$\displaystyle B_{a\,i}=\frac{1}{2}\epsilon_{ijk}F_a^{jk}$
and $\lambda^a$ are $n$ Lagrangian multipliers
identical to the first component of the gauge potentials $\lambda^a=A^a_{\scriptscriptstyle 0}$. 

In the 50's Dirac \cite{Dirac} recognized the necessity of a gauge-fixing procedure, already at the classical 
level, when he tried to give an hamiltonian formulation for a system with first class constraints.
This necessity is particularly evident in the CPI approach. In fact from the Hamiltonian (\ref{ma}) 
we can derive the following equations of motion:
\begin{eqnarray}
\lefteqn{\dot{\pi}^k_a=-\partial_iF^{ki}_a-\lambda^dC^b_{da}\pi^k_b+C^b_{ac}A^c_iF^{ki}_b}\nonumber\\
\lefteqn{\dot{A}^a_k=\pi^a_k+\partial_k\lambda^a+C^a_{dc}\lambda^dA^c_k} \label{eqq}
\end{eqnarray}
Let us indicate with: $\phi^{\scriptscriptstyle A}=(A^a_k,\pi^k_a)$ both the spatial components of the 
fields and the momenta. We can then rewrite the equations of motion in terms of a
$6n\times 6n$ antisymmetric matrix $\omega^{\scriptscriptstyle AB}$:
$\displaystyle \dot{\phi}^{\scriptscriptstyle A}=\omega^{
\scriptscriptstyle AB}\frac{\partial H}{\partial \phi^{\scriptscriptstyle B}}$
where we indicate with $\partial$ the functional derivative and $H$ is given by eq. (\ref{ma}). 
At this point in the CPI procedure the step $
\displaystyle {\widetilde{\delta}}(\phi^{\scriptscriptstyle A}-\phi^{\scriptscriptstyle A}_{cl})\rightarrow{\widetilde{\delta}}
(\dot{\phi}^{\scriptscriptstyle A}-\omega^{\scriptscriptstyle AB}\frac{\partial H}{\partial \phi^{\scriptscriptstyle B}})$
is {\it forbidden} since,  because of the arbitrariness of $\lambda^a$, 
there is not a one-to-one correspondence between solutions and equations
of motion. 
So we have to {\it fix the gauge}. 
We can, for example, enforce the Lorentz-gauge condition by
inserting into the original path integral:
\begin{equation}
Z_{\scriptscriptstyle CPI}^{\scriptscriptstyle YM}=\int {\cal D}\phi \;{\widetilde{\delta}}(\phi^{\scriptscriptstyle A}-\phi
^{\scriptscriptstyle A}_{cl}) \label{tre}
\end{equation}
the following expression formally indipendent of the fields:
\begin{equation}
1=\int {\cal D}A^{a}_0\;{\widetilde{\delta}}
(-\partial^{\mu}A^{a}_{\mu})\Delta_F[{\bf A}]	      \label{cinque}
\end{equation}
where $\Delta_F[{\bf A}]\sim det[-\partial^{\mu}\delta^{a}_cD^{bc}_{\mu}]$ is the usual Faddeev-Popov
determinant.
Having fixed the gauge, we can now perform the transformation from 
the solutions to the equations of motion in (\ref{tre}):
\begin{eqnarray}
\lefteqn{\widetilde{\delta}(\phi^{\scriptscriptstyle A}-\phi^{\scriptscriptstyle A}_{cl})={\widetilde{\delta}}
(\dot{\phi}^{\scriptscriptstyle A}-w^{\scriptscriptstyle AB}\partial_{\scriptscriptstyle B}H)\cdot}\label{quattro} \\
&&\qquad\qquad\cdot det(\delta_
{\scriptscriptstyle B}^{\scriptscriptstyle A}\partial_t
-w^{\scriptscriptstyle AC}\partial_{\scriptscriptstyle B}\partial_{\scriptscriptstyle C}H)\nonumber  
\end{eqnarray}
where all the derivatives have to be intended as functional ones.
Next we have to exponentiate not only the RHS of (\ref{quattro}) but also the gauge-fixing part (\ref{cinque}) in order 
to give an 
exponential weight to the CPI. We will name this CPI version of YM theories as NAT (natural) 
because this is the most natural and direct way of fixing the gauge:
\begin{equation}
Z^{\scriptscriptstyle NAT}_{\scriptscriptstyle CPI}=\int {\cal D}{\mu}\; exp\;i\int d^4x
\widetilde{\cal L}^{\scriptscriptstyle NAT} \label{natural}
\end{equation}
The functional integration is extended over all the variables $\mu$ of the theory (see also the next section)
and the Lagrangian density is:
\begin{eqnarray} 
\lefteqn{\displaystyle \widetilde{\cal L}^{\scriptscriptstyle NAT}=-\pi_{a}\partial^{\mu}A^{a}_{\mu}
+\Lambda_{\scriptscriptstyle A}(\dot{\phi}^{\scriptscriptstyle A}-
\omega^{\scriptscriptstyle AB}\partial_{\scriptscriptstyle B}H)} \label{lagnat} \\
\lefteqn{-i\partial^{\mu}\bar{C}_aD^{\;\;a}_{b\mu}C^b
+i\bar{\Gamma}_{\scriptscriptstyle A}(\delta^{\scriptscriptstyle A}_{\scriptscriptstyle B}\partial_t-
\omega^{\scriptscriptstyle AC}\partial_{\scriptscriptstyle B}
\partial_{\scriptscriptstyle C}H)\Gamma^{\scriptscriptstyle B}} \nonumber
\end{eqnarray}
In (\ref{lagnat}) $\pi_a$ is the variable we use to exponentiate the gauge fixing condition,
$\bar{C}_a$ and $C^b$ are the usual Faddeev-Popov ghosts, while the $(\Lambda,\Gamma,\bar{\Gamma})$ are the analogue
of the $(\lambda,c,\bar{c})$ of the CPI of the point particle \cite{Gozzi1}.

\section{BFV method}

In quantum field theory there is also a more general way of implementing the gauge-fixing procedure:
the Batalin, Fradkin and Vilkovisky (BFV) method. In this approach one enlarges the original phase-space
$(A^a_k,\pi^k_a)$ to include as dynamical variables also the Lagrangian multipliers $\lambda^a$, their
conjugate momenta $\pi_a$, and a number of BFV ghosts $\eta^a=(-iP^b,C^b)$ and antighosts 
${\cal{P}}_a=(i\bar{C}_b,\bar{P}_b)$ equal to the number of constraints $\psi_a=(\pi_b,\sigma_b)$
present in the theory. All the ghosts and the constraints contribute in building the following
BRS-BFV charge:
\begin{equation}
\Omega^{\scriptscriptstyle BFV}=\int
d^3x [\sigma_aC^{a}-iP^{a}\pi_a+\frac{1}{2}\bar{P}_aC^{a}_{bc}C^bC^c]
\end{equation}
For reasons that are clearly
explained in \cite{Henneaux1}-\cite{Henneaux2}, we have to identify two Hamiltonian densities that 
differ for a BRS-exact term:
${\cal H}^{\scriptscriptstyle BFV}={\cal H}_{\scriptscriptstyle YM}-\{\theta,\Omega^{\scriptscriptstyle BFV}\}$, 
where the $\{\cdot,\cdot\}$ are the graded Poisson brackets that 
one can introduce in the BFV phase-space. In this case fixing the gauge means choosing a gauge
function $\theta$. If we choose:
$\theta=i\bar{C}_a\partial^kA^{a}_k+\bar{P}_a\lambda^{a}$ then we obtain the following Hamiltonian density:
\begin{eqnarray}
\displaystyle 
\lefteqn{{\cal H}^{\scriptscriptstyle BFV}=\frac{1}{2}\pi^k_a\pi^{a}_k+\frac{1}{4}F_a^{ij}F^{a}_{ij}
+\pi_a\partial^kA^{a}_k}\nonumber\\
\lefteqn{\qquad\;\;\;-\lambda^{a}\partial_k\pi^k_a+\lambda^{a}C^b_{\;ac}
\pi^k_bA^c_k
+i\bar{P}_aP^{a}}\\
\lefteqn{\qquad\;\;\;-\lambda^{a}\bar{P}_bC^b_{\;ac}C^c-i\bar{C}_a\partial^k(\partial_kC^{a}+C^{a}_{\;bc}A^c_kC^b)}\nonumber
 \label{hamiltonian}
\end{eqnarray} 
As the BFV method has automatically managed to implement the gauge-fixing procedure, 
we can then proceed to build the CPI
without adding any further ingredient.
First, we can derive the equations of motion and put them in the usual symplectic form:
$\displaystyle  \dot{\xi}^{\scriptscriptstyle  A}=\omega^{\scriptscriptstyle AB}\vec{\partial}_{\scriptscriptstyle B}
H^{\scriptscriptstyle BFV}$.
We indicate with $\xi^{\scriptscriptstyle A}$ all the fields, including the gauge ghosts, 
present in $H^{\scriptscriptstyle BFV}=\int d^3x {\cal H}^{\scriptscriptstyle BFV}$. 
So the grassmannian character of the fields is mixed: there are grassmannian even as well as
odd fields. This will cause some formal complications. For example to pass from the solutions
to the equations of motion we need an object that is known in literature as the superdeterminant or
berezinian. The superdeterminant of a general supermatrix $M^{\scriptscriptstyle A}_{\scriptscriptstyle B}$
can be exponentiated \cite{Carta} using auxiliary variables which have 
a grassmannian parity opposite to that of the fields they refer to:
\begin{equation}
\displaystyle sdet(M^{\scriptscriptstyle A}_{\scriptscriptstyle B})=
\int{\cal D}\bar{\Gamma}{\cal D}\Gamma
exp-\int d^4x \bar{\Gamma}_{\scriptscriptstyle A}M^{\scriptscriptstyle A}_{\scriptscriptstyle B}\Gamma^{
\scriptscriptstyle B}
\end{equation}
where $[\Gamma^{\scriptscriptstyle A}]\equiv[\Gamma^{\xi^A}]=[\bar{\Gamma}_{\xi^A}]=[\xi^{\scriptscriptstyle A}]+1$
(we indicate in square brackets
the grassmannian parity). 
In the same way the exponentiation of the equations of motion can be done using suitable
auxiliary variables $\Lambda_{\scriptscriptstyle A}$, which have the same grassmannian parity with respect to 
the fields they refer to: $[\Lambda_{\scriptscriptstyle A}]\equiv[\Lambda_{\xi^A}]=[\xi^{\scriptscriptstyle A}]$.
In this way we can construct the following CPI for YM theory \cite{Carta}:
\begin{equation}
\displaystyle Z^{\scriptscriptstyle BFV}_{\scriptscriptstyle CPI}=\int {\cal D} 
\Lambda_{\scriptscriptstyle A} {\cal D} \xi^{\scriptscriptstyle A} {\cal D} \bar{\Gamma}_
{\scriptscriptstyle A} {\cal D}
\Gamma^{\scriptscriptstyle A} e^{ i \int d^4x \widetilde{\cal L}^{\scriptscriptstyle BFV}} \label{second}
\end{equation}
where:
\begin{equation}
\widetilde{\cal L}^{\scriptscriptstyle BFV}= \Lambda_{\scriptscriptstyle A}\dot{\xi}^A
+i\bar{\Gamma}_{\scriptscriptstyle A}\dot{\Gamma}^{\scriptscriptstyle A}-\widetilde{\cal H}^{\scriptscriptstyle BFV}
\end{equation}
and:
\begin{equation}
\widetilde{\cal H}^{\scriptscriptstyle BFV}=\Lambda_{\scriptscriptstyle A} \omega^{\scriptscriptstyle AB}\vec{\partial}_
{\scriptscriptstyle B}H
+i\bar{\Gamma}_{\scriptscriptstyle A}\omega^{\scriptscriptstyle AC}\stackrel{\rightarrow}{\partial}
_{\scriptscriptstyle C}H\stackrel{\leftarrow}{\partial}_{\scriptscriptstyle B}\Gamma^{\scriptscriptstyle B} \label{CPIham}
\end{equation}
Somehow we have constructed {\it two} different classical path integrals for Yang-Mills theories:
the natural one, $\displaystyle Z_{\scriptscriptstyle CPI}^{\scriptscriptstyle NAT}$ of eq. (\ref{natural}),
and the BFV one, $\displaystyle Z_{\scriptscriptstyle CPI}^{\scriptscriptstyle BFV}$ of eq. (\ref{second}). 
The functional integrations entering the 
$\displaystyle Z_{\scriptscriptstyle CPI}^{\scriptscriptstyle NAT}$ and
$\displaystyle Z_{\scriptscriptstyle CPI}^{\scriptscriptstyle BFV}$ are over a different number of fields. 
We have proved in \cite{Mauro} that several integrations in 
$\displaystyle Z_{\scriptscriptstyle CPI}^{\scriptscriptstyle BFV}$ can be done explicitly
to bring:
$\displaystyle Z_{\scriptscriptstyle CPI}^{\scriptscriptstyle BFV}=
\int{\cal{D}}\mu^{\prime} \; exp \;i\int d^4x\tilde{\cal{L}}^{\scriptscriptstyle BFV}$
down to:
$\displaystyle Z_{\scriptscriptstyle CPI}^{\scriptscriptstyle NAT}=
\int{\cal{D}}\mu \; exp \;i\int d^4x\widetilde{\cal{L}}^{\scriptscriptstyle NAT}$.
In a certain sense we can say that this is the classical counterpart of the proof of the equivalence of the Faddeev-Popov
and BFV methods in quantum field theory \cite{Henneaux1}.

\section{CPI superfields in YM theories}
In the following table we will write down all the fields of the CPI-BFV theory (\ref{second}); 
among all these fields
we shall indicate with bold characters those ones which do not enter the $\displaystyle Z_{\scriptscriptstyle CPI}^{\scriptscriptstyle NAT}$:
 \begin{tabbing}
 $\mathbf{\bar{\Gamma}_{A^a_k}}$ \= $\mathbf{\bar{\Gamma}_{\pi^k_a}}$ \= $\bar{\Gamma}_{\lambda_a}$ \= $\bar{\Gamma}_{\pi_a}$ \=
 $\bar{\Gamma}_{C^a}$ \= $\bar{\Gamma}_{\bar{C}_a}$ \= $\bar{\Gamma}_{P_a}$ \= $\bar{\Gamma}_{\bar{P}_a}$  \kill 
 ${A^a_k}$ \> ${\pi^k_a}$ \> ${\lambda^a}$ \> ${\pi_a}$ \>
 ${C^a}$ \> ${\bar{C}_a}$ \> $\mathbf{P_a}$ \> $\mathbf{\bar{P}_a}$ \\
 ${\Gamma^{A^a_k}}$ \> ${\Gamma^{\pi^k_a}}$ \> $\mathbf{\Gamma^{\lambda_a}}$ \> $\mathbf{\Gamma^{\pi_a}}$ \>
 $\mathbf{\Gamma^{C^a}}$ \> $\mathbf{\Gamma^{\bar{C}_a}}$ \> $\mathbf{\Gamma^{P_a}}$ \> $\mathbf{\Gamma^{\bar{P}_a}}$ \\
 ${\bar{\Gamma}_{A^a_k}}$ \> ${\bar{\Gamma}_{\pi^k_a}}$ \> $\mathbf{\bar{\Gamma}_{\lambda_a}}$ \> 
 $\mathbf{\bar{\Gamma}_{\pi_a}}$ \>
 $\mathbf{\bar{\Gamma}_{C^a}}$ \> $\mathbf{\bar{\Gamma}_{\bar{C}_a}}$ \> $\mathbf{\bar{\Gamma}_{P_a}}$ \> 
 $\mathbf{\bar{\Gamma}_{\bar{P}_a}}$ \\
 ${\bar{\Lambda}_{A^a_k}}$ \> ${\bar{\Lambda}_{\pi^k_a}}$ \> $\mathbf{\bar{\Lambda}_{\lambda_a}}$ \> 
 $\mathbf{\bar{\Lambda}_{\pi_a}}$ \>
 $\mathbf{\bar{\Lambda}_{C^a}}$ \> $\mathbf{\bar{\Lambda}_{\bar{C}_a}}$ \> $\mathbf{\bar{\Lambda}_{P_a}}$ \> 
 $\mathbf{\bar{\Lambda}_{\bar{P}_a}}$
 \end{tabbing}
The fields that appear in the same vertical line in the previous table can be put together and considered
as the components of a multiplet which was called the CPI superfield in ref. \cite{Gozzi1}. 
With respect to the point particle case we need to introduce two different superfields:
an {\it even} superfield defined as:
\begin{equation}
\Xi^{a}=\xi^{a}+\theta\Gamma^{a}+\bar{\theta}\omega^{ab}\bar{\Gamma}_b
+i\bar{\theta}\theta\omega^{ab}\Lambda_b
\end{equation}
and an {\it odd} one defined as:
\begin{equation}
\Xi^{\alpha}=\xi^{\alpha}-\theta\Gamma^{\alpha}-\bar{\theta}\omega^{\alpha\beta}\bar{\Gamma}_\beta
-i\bar{\theta}\theta\omega^{\alpha\beta}\Lambda_\beta
\end{equation}
Notice that they differ for some signs.
We can then easily show that the ${\cal H}^{\scriptscriptstyle BFV}$ and the 
associated $\widetilde{\cal H}^{\scriptscriptstyle BFV}$ are related as follows:
\begin{equation}
i\int d\theta d\bar{\theta}\; {\cal H}^{\scriptscriptstyle BFV}(\Xi)=\widetilde{\cal{H}}^{\scriptscriptstyle BFV}
\end{equation}
The concept of superfield allows us also to easily map the quantum simmetry charges into the
corresponding ones of the CPI. For example, from $\Omega^{\scriptscriptstyle BFV}$ we can construct
the correspondent $\widetilde{\Omega}^{\scriptscriptstyle BFV}$ charge as:
\begin{equation}
i\int d\theta d\bar{\theta} \; \Omega^{\scriptscriptstyle BFV}(\Xi)=\widetilde{\Omega}^{\scriptscriptstyle BFV}
\end{equation}
It is possible to prove that the conservation of
$\Omega^{\scriptscriptstyle BFV}$ under ${\cal H}^{\scriptscriptstyle BFV}$ implies the conservation of
$\widetilde{\Omega}^{\scriptscriptstyle BFV}$ under the evolution generated by 
$\widetilde{\cal H}^{\scriptscriptstyle BFV}$. 
Similarly to what happens in the quantum case, the classical action 
$\widetilde{S}^{\scriptscriptstyle BFV}=\int d^4x \widetilde{\cal L}^{\scriptscriptstyle BFV}$ 
is invariant under the transformations generated by $\widetilde{\Omega}^{\scriptscriptstyle BFV}$.
This property is crucial to {\it guarantee the gauge-invariance} of the physical results at the classical level.

The {\it universal} charges of the CPI found for the point particle
can be found also in the BFV version of the YM-CPI.
For example the BRS charge of the CPI becomes:
\begin{equation}
\displaystyle Q^{\scriptscriptstyle BRS}_{\scriptscriptstyle CPI}
=i\int d^3x \Gamma^{\scriptscriptstyle A}\Lambda_{\scriptscriptstyle A}
\end{equation}
The physical meaning of $\displaystyle Q^{\scriptscriptstyle BRS}_{\scriptscriptstyle CPI}$ and 
$\displaystyle \widetilde{\Omega}^{\scriptscriptstyle BFV}$ 
is very different. In fact, if we return to the table of all the fields of the
BFV method, we have that
the action of 
$\displaystyle \widetilde{\Omega}^{\scriptscriptstyle BFV}$ on the fields of the first row 
is identical to that of the quantum charge $\Omega^{\scriptscriptstyle BFV}$. 
They both generate the gauge transformations of the theory and
mix the fields horizontally. Anyhow, differently than $\Omega^{\scriptscriptstyle BFV}$, the 
$\widetilde{\Omega}^{\scriptscriptstyle BFV}$ acts also on the fields of the the other rows mixing them. 
The $Q^{\scriptscriptstyle BRS}_{\scriptscriptstyle CPI}$ charge, instead, 
turns the fields of the first row into the associated Jacobi fields contained in the second row and so it allows us to
move vertically through the table. Even if we called both charges as BRS charges they actually perform
different operations.

The reader may ask which is the goal of all the machinery we have built. What we have in mind 
is a geometrical goal. In fact in ref. \cite{Topologico} it was proved that, choosing proper boundary conditions,
the CPI of the point particle can be turned into a topological field theory and used to calculate, among other things,
the Euler number of the manifold on which the system lives. Performing the same thing with 
the $Z_{\scriptscriptstyle CPI}^{\scriptscriptstyle YM}$ we could in principle calculate the analog of the
Euler number for the space of gauge orbits or other geometrical characteristics of this space. All these geometrical
features are crucial for a better grasp of the non perturbative regime of the YM theories.

The authors would like to thank E. Gozzi for many helpful discussions and suggestions.    
This research was supported by grants from INFN, MURST and the Univ. of Cagliari and Trieste.

\end{document}